\documentstyle[aps,preprint]{revtex}
\tightenlines
\begin{document}
\draft
\title{
Magnetic field effects on two-dimensional Kagome lattices
}
\author{
Takashi Kimura,${}^1$ Hiroyuki Tamura,${}^1$ Kenji Shiraishi,${}^2$ and Hideaki Takayanagi${}^1$
}
\address{
${}^1$NTT Basic Research Laboratories, Morinosato-Wakamiya 3-1, Atsugi 243-0198, Japan
${}^2$Department of Physics, Tsukuba University, Tsukuba 305-8271, Japan
}
\date{\today}
\maketitle
\begin{abstract}
 Magnetic field effects on single-particle energy bands (Hofstadter butterfly), Hall 
conductance, {\it flat-band ferromagnetism}, and magnetoresistance of two-dimensional 
{\it Kagome lattices} are studied. 
The flat-band ferromagnetism is shown to be broken as the 
flat-band has finite dispersion in the magnetic field. A metal-insulator transition 
induced by the magnetic field ({\it giant negative magnetoresistance}) is predicted. In the 
half-filled flat band, the ferromagnetic-paramagnetic transition and the metal-insulator 
one occur simultaneously at a magnetic field for strongly interacting electrons.
All of the important magnetic fields effects should be observable 
in mesoscopic systems such as quantum dot superlattices. 
\end{abstract}
For the last two decades, low-dimensional systems have been one of the most important 
subjects of research in condensed matter physics. Especially, two-dimensional (2D) 
electron systems have been of great interest because of striking phenomena such as 
high-Tc superconductivity,  and the (fractional) quantum Hall effect (QHE).
For the last decade, the ferromagnetism of 2D systems 
has also attracted attention, especially for systems without 
any magnetic elements. The ferromagnetism of the 2D electron gas has 
never been observed because the parameter, which is the ratio of the mean distance 
between nearest electrons to the Bohr radius, must become very large ($\sim 40$ ) for the 
ferromagnetism \cite{Tanatar1989} even if we neglect the Wigner crystallization. However, the 
ferromagnetism has been rigorously proved for Hubbard models on some lattice 
structures having a {\it flat band} (FB) 
\cite{Lieb1989,Mielke1991,Tasaki1992}. The {\it flat-band ferromagnetism} (FBF) is 
not only expected as a starting point for understanding the ferromagnetism of itinerant 
electrons but also for fabrication to make a FB 
in real materials \cite{Shima1993,Fujita1996,Yajima1999}. 
At present, however, there is no clear evidence of the FBF in real materials because 
of difficulties such as uncontrollable electron filling 
and the Jahn-Teller effect that breaks the 
FB degeneracy. Recently, 
we have proposed a method of making an effective Hubbard 
model with some bipartite lattices \cite{Tamura2000} and 
the {\it Kagome lattice} structure \cite{Shiraishi2001}, 
which have a FB, in mesoscopic quantum dot superlattices. 
For dot superlattices, 
we can avoid the above-mentioned problems in bulk systems,    
and the parameters of the Hubbard model, the electron filling, 
and the lattice structure can be freely changed. 
Furthermore, there is another important advantage 
of a mesoscopic system when we 
study magnetic field effects. Of course, there are lots of important magnetic 
field effects in bulk systems. However, in dot superlattices it is easy to study magnetic field 
effects at the {\it lattice level}, 
such as the Hofstadter butterfly and lattice QHE\cite{Albrecht2001}. In bulk systems 
the magnetic field that is needed in order to input a magnetic-flux quantum within a unit cell, 
which has a lattice constant of few angstroms, is of the order of $10^4$ T, 
but in dot superlattices, 
the magnetic field needed is very small and realistic. For example, 
in Kagome dot superlattices proposed in Ref. \cite{Shiraishi2001},  
the lattice constant is of the order of 100 nm, and the magnetic field needed is of the  
order of 0.1 T. In addition, if we pay attention to the many-body effects of electrons, 
it is quite interesting to study the magnetic field effects on the FBF of the Kagome lattice. 

In this Letter, we study magnetic field effects on the Kagome lattice. In the magnetic 
filed, the FB is broken down because it originates from the interference of the 
phases of wave functions. 
Hall conductance at a magnetic field is shown to have a wide 
Hall plateau, which may be easily observable in experiments.
The FBF, which has been mathematically 
proved by Mielke \cite{Mielke1991}, is also broken by 
the magnetic field, which is understood to be a result of competition between generalized 
Hund's coupling and the single-particle energy. This ferromagnetic-paramagnetic 
transition is a new magnetic field effect. Usually, the magnetic field  
supports ferromagnetism by Zeeman coupling, which favors the aligned spins along 
the direction of the magnetic field. We should note that in dot superlattices the Zeeman effects 
are negligible for small magnetic field. We also find a {\it giant negative 
magnetoresistance} (GNMR), which is the metal-insulator transition induced by the 
magnetic field that breaks the FB. Furthermore, when the FB is half-filled, the 
ferromagnetic-paramagnetic transition and the metal-insulator one occur simultaneously 
for strongly interacting electrons in a magnetic field. \par 

Let us start from the single-particle properties of the Kagome lattice in a magnetic field. 
We assume a tight-binding model for the Kagome lattice [Fig. 1(a)]. The magnetic field is 
incorporated in hopping integral $t_{ij}$  in the usual manner through the Peierls phase for a 
gauge field as
\begin{eqnarray}
t_{ij}(\bf{A}) \equiv \it{t} {\rm exp} (\rm{i}\theta_{ij}),\hspace{0.2cm}
\theta_{ij}=-\frac{2\pi}{\phi_0}\int^{\bf{r}_j}_{\bf{r}_i}{\bf A}\cdot d\bf{r},
\end{eqnarray}
where $\bf{A}$  is a vector potential for the magnetic field $\phi$, $\phi_0\equiv h/e$ is 
the magnetic-flux quantum, and $t (>0)$ is the hopping integral 
at zero magnetic field. Hereafter, we use a unit where a magnetic-flux quantum through the
smallest triangle in a unit cell is defined as unity. Our result for the Hofstadter butterfly 
(the single-particle energy spectrum) of the Kagome lattice is shown in Fig 1(b). 
When there is no magnetic field, 
there is a FB at the $E/t=2$ [Fig. (2)] and it is broken by applying magnetic 
field. The magnetic field changes the phase of the wave function and the interference 
between the electron wave functions. As a result, the magnetic field breaks the 
interference-originated FB in the Kagome lattice. We note that the FB is not always 
broken by a magnetic field for other lattice structures. For example, the FB of the Lieb 
lattice is preserved even in a magnetic field \cite{Aoki1996}. 
The difference is due to the origins of the FBs. 
In the Lieb lattice the FB originates not from the interference but from the lattice 
topology. Let us revert to the subject of the Kagome lattice.
In the Kagome lattice, for $\phi=n/8m$ ($n$, $m$: integer), 
there are $3m$ magnetic bands, and for $\phi=n/8$ the number of the magnetic bands 
is smallest and the band gap is widest. 
In real systems it is indeed difficult to observe the 
physical properties of the small band-gap 
structures of the Hofstadter butterfly because of the inevitable 
decoherence of the wave functions. Hence only the large band-gap structures, such as $\phi=n/8$,
will be experimentally important. On the other hand, it is well known that the quantized 
Hall conductance reflects the band-gap structures directly. We calculated the Hall 
conductance, which is equivalent to the Hall conductivity in 2D, 
with the Kubo formula \cite{Thouless1982,Kohmoto1985}. Figure 3 shows the 
Hall conductance as a function of Fermi energy for $\phi=1/8$ and $1/4$. 
Especially for $\phi=1/4$, we can see a wide 
quantized Hall plateau reflecting a wide band gap: there is indeed a flat band at $E/t=0$,
but its contribution to the conductance is zero and the plateau is not changed there. The QHE may be observable in experiments without great effort 
because of the wide plateau. 
We neglected the interaction between 
electrons in the above calculations. But if we assume the interaction is not so strong and 
short-ranged, such as that in quantum dot superlattices, 
it may not significantly affect the Hall 
plateau. \par

We must consider the interaction between electrons to study the FBF. Let us consider the 
Hubbard model with the on-site interaction $U$ between electrons with opposite spins. 
The Hamiltonian is written as
\begin{eqnarray}
H=-\sum_{<i,j>}(t_{ij}({\bf A})c_{i\sigma}^\dagger c_{j\sigma}+\rm{h.c.})
+\it{U\sum_{i}n_{i\uparrow}n_{i\downarrow}},
\end{eqnarray}
where  $n_{i\sigma}\equiv c_{i\sigma}^\dagger c_{i\sigma}$, $c_{i\sigma}$ 
annihilates an electron with spin $\sigma$ at site $i$, and $<\cdot\cdot\cdot>$ refers to pairs 
of nearest neighbors. For the Kagome Hubbard lattice without a magnetic field, Mielke 
\cite{Mielke1991} has mathematically 
proved the ferromagnetic behavior in a finite region where the 
filling of the FB equals or is slightly larger than the half-filling. 
Reference \cite{Tamura2001} has 
studied the Kagome Hubbard cluster without a magnetic field. There we can see the 
highest-spin state for the half-filled FB, and the height of the total spin is gradually 
decreased for larger fillings as expected. \par
We study the magnetic field effects on the Kagome Hubbard clusters. We use the 
numerical exact-diagonalization method with the Lanzos algorithm \cite{Dagotto1994}. We 
employed the anti-periodic boundary condition to avoid excess degeneracy at the $\Gamma$
point to reduce the finite-size effects (see Fig. 2). Our results for the magnetic field 
effect on the FBF of the 12-site (2x2 unit cells) cluster are shown in Table 1, where the 
total spins of the ground states are presented for the filling that equals or is slightly 
larger than half-filled where Mielke's proof can be applied for the FBF at zero 
magnetic fields. We can see the ferromagnetic-paramagnetic transition induced by the 
magnetic field. The origin of the transition is the breakdown of the FB that 
produces the FBF. The FBF originates from a generalized Hund's coupling, which 
results in a high-spin state. 
One may imagine that in a FB there exist localized states that are 
mutually disjointed. Remarkably, this does not hold for the FB of the Kagome lattice. 
Here the most compact states cannot be confined within the unit cell and must be 
overlapped with each other. The overlap between the compact states is advantageous for 
the high-spin states that exploit Pauli's principle to avoid the repulsive interaction 
between electrons with opposite spins. However, if the FB is broken by the magnetic 
field, the single-particle energy of the high-spin states becomes larger than that of the 
low-spin state and the FBF is also broken by large magnetic fields. Here we note that 
the results for 3x2 unit cells are qualitatively the same as those for 2x2 unit cells. \par

 Finally, we discuss the magnetoresistance when the FB crosses the Fermi level, 
calculating the Drude weight $D$, which is defined as the zero frequency part of the optical 
conductivity $\sigma(\omega)\equiv D\delta(\omega)+ \rm{regular\ part\ for\ finite}\ \omega$. 
The Drude weight is given by \cite{Dagotto1994}
\begin{eqnarray}
\frac{D}{2\pi e^2}=-\frac{\langle\Psi_0 |\hat{K}|\Psi_0\rangle}{2N_s}-
\frac{1}{N_s}\sum_{n\neq 0}\frac{|\langle\Psi_0 |\hat{J}|\Psi_0\rangle|^2}{E_n-E_0}\ .
\end{eqnarray}
Here $\Psi_{0(n)}$ is the ground (n-th excited) state with energy eigenvalue $E_{0(n)}$, $N_s$  
is the number of sites, 
${\hat{J}}$ is the current operator, and $\hat{K}$  is the "kinetic energy" 
operator along the direction of the current. In the bulk limit, generally speaking, if $D=0$,
the system is insulating, while if $D>0$ it is metallic (or superconducting). We present our 
results for the Drude weights of the ground states for Kagome clusters with 2x2 unit cells 
in the usual manner \cite{Dagotto1994,notice1}: 
we also calculated the Drude weights for 3x2 unit 
cells and obtained qualitatively the same results as those for 2x2 unit cells. Figure 4 shows 
the magnetic field dependence of the Drude weight. Let us start from the case when the 
filling is away from half-filled on the FB. All data of the Drude weight are 
overlapped at $D=0$, indicating the insulating behavior at zero magnetic field, and 
become finite, indicating the metallic behavior when the magnetic field is applied [Fig. 
4(a)]. The metal-insulator transition induced by the magnetic field leads to the GNMR 
of the Kagome lattice. The GNMR is clearly understood within a single-particle picture. 
The group velocity of electrons is zero without the magnetic field and the system is 
insulating. In magnetic field the FB has a finite curvature and the group velocity 
becomes finite, resulting in a metallic system. \par

We also calculated the Drude weights at the half filled FB [Fig. 4(b)]. For small $U/t<3.8$ 
[data for $U/t=0,3$ in Fig. 4(b)] the magnetic field dependence is qualitatively the same as 
that for the case away from half filling. However, for large $U/t$  [data for 
$U/t=5,50,100$ in Fig. 4(b)], we find the metal-insulator transition occurs simultaneously 
with the ferromagnetic-paramagnetic one for a magnetic field ($\phi=1/8$). We can see for 
large $U/t$ that the high-spin ($S=2$) and 
insulating states with negligible Drude weights below 
$\phi=1/8$ turn into the low-spin ($S=0$) and metallic states with finite Drude weights above 
$\phi=1/4$: at $\phi=1/8$ the data for 
$U/t=5,50,100$ are almost completely overlapped at $D=0$. 
Let us consider the physical picture for these simultaneous transitions. The electrons on 
the FB for the high-spin state have the same spin direction (defined as the up direction 
without any loss of generality). We should remember that the number of $k$-points and 
that of the electrons are exactly the same at half filling. Thus, all the up-spin states for 
the high-spin states are completely occupied at the half filling, while all the down-spin 
states are unoccupied. Thus, the up-spin electrons on the FB cannot move there because 
of Pauli's exclusion principle and the absence of any spin-flipping processes in the 
present Hamiltonian. These are the reasons the high-spin states are insulating at half 
filling. Above $\phi=1/4$, the metallic states appear because there the ground states turn to 
low-spin ones where both unoccupied and occupied states co-exist for both up and 
down spins and electrons can now move, resulting in the simultaneous 
ferromagnetic-paramagnetic and metal-insulator transitions. On the other hand, we 
also note that the Drude weight is converged to a finite value for the large $U/t$ limit for   
low-spin states: even above $\phi=1/4$, we indeed find the high-spin ground states for very 
large $U/t$ (for e.g. $U/t=110$ at $\phi=1/4$). 
Thus there may be no Mott transition for low-spin 
states in the usual sense even though the FB is half-filled. This is because the 
half-filling of the FB does not correspond to 
half-filling of the total system having multi-bands. 
This means that electrons can move, avoiding additional on-site interaction 
energy in real space. This situation is different from single-band Hubbard models at half 
filling. There the electron number per one site is exactly one and the interaction energy 
(of the order of $U$) is needed for one electron to move from the first site to its nearest 
neighbor. \par

In conclusion, we have studied magnetic field effects on a 2D Kagome lattice. 
The Hofstadter butterfly shows wide gap structures that may be easily 
observable in real experiments 
through the wide plateaus of the QHE. On the flat band, the 
ferromagnetic-paramagnetic and metal-insulator transitions (GNMR) induced by the 
magnetic field are predicted. At the half-filled FB, we find the metal-insulator transition 
occurs simultaneously with the ferromagnetic-paramagnetic one for large $U/t$ at a 
magnetic field. 
All of the above magnetic field effects should be observable for small 
(order of 0.1T) magnetic fields in quantum dot superlattices. \par

  We thank Prof. K. Kuroki for fruitful discussions. We also thank Dr. S. 
Ishihara for his continuous encouragement and helpful advice.
This work was partly supported by the NEDO International Joint Research Grant.

\begin{figure}
\caption{
The Kagome lattice structure and the magnetic flux though the Kagome lattice is 
schematically shown in Fig. (a). Figure (b) shows the single-particle energy levels in the 
magnetic field (i.e. Hofstadter butterfly). 
}
\label{fig1}
\end{figure}

\par

\begin{figure}
\caption{
Single-particle band structure of the Kagome lattice without a magnetic field
is shown.
}
\label{fig2}
\end{figure}
\begin{figure}
\caption{
Hall conductance $\sigma_{xy}$ is shown for both $\phi=1/8$ (a) and $\phi=1/4$ (b) as a 
function of Fermi energy. 
}
\label{fig3}
\end{figure}
\begin{figure}
\caption{
Magnetic field dependences of the Drude weights of 12-site (2x2 unit cells) 
Kagome Hubbard cluster are shown. Figure (a) shows the results when the filling is away from 
half filling of the FB, while Fig. (b) shows those at half filling. 
}
\label{fig4}
\end{figure}



\begin{table}

\caption{
Total spins of the ground state of the 12-site (2x2 unit cells) Kagome Hubbard 
cluster with $U/t=5$ are shown as a function of the electron number $N$ and magnetic field $\phi$. 
}
\label{table4}
\end{table}


\begin{references}
\bibitem{notice1}  Exactly speaking, 
the Drude weight slightly depends on the direction of the current 
not qualitatively but quantitatively. We present only the Drude weights for the direction 
that is perpendicular to a side of the smallest triangle in the unit cell.
\end{references}
\end{document}